# A Performance Study on the Throughput and Latency of Zenoh, MQTT, Kafka, and DDS

**Wen-Yew Liang**[*]
william.l@zettascale.tech

**Yuyuan Yuan**[+]
yyyuanowo@gmail.com

**Hsiang-Jui Lin**[+]
jerry73204@gmail.com

[*] Zenoh Taiwan Team, Zettascale Technology
[+] Department of Computer Science and Information Technology, National Taiwan University

## Abstract

In this study, we compare the performance of the new-generation communication protocol Zenoh with the widely-used MQTT, Kafka, and DDS. Two performance indexes were evaluated, including throughput and latency. A brief description of each protocol is introduced in this article. The experiment configuration and the testing scenarios are described in detail. The results show that Zenoh outperforms the others with impressive performance numbers.

## Introduction

As the requirement for shorter response time for applications like IIoT is increasing, the traditional cloud-based computing model is now moving its focus to the edge of the network. To support that, topic-based publication/subscription (a.k.a. pub/sub, with push or pull models) has been thought to be the most widely-used communication method, with several implementations provided for field applications. Among them, MQTT and Kafka have been considered the most popular ones. Another well-established pub/sub standard is Data Distribution Service (DDS) [1]. DDS is an OMG (Object Management Group) standard for high-performance, highly reliable, and decentralized real-time connectivity, which has been widely used in military, aerospace, and transportation applications.

Zenoh [2] is a newly-developed pub/sub-based data-centric communication protocol supporting cloud-to-edge and things continuum with the characteristics of high performance and high scalability and is highly reliable. It relies on a fully decentralized architecture across all kinds of computation platforms and is designed to perfectly fit into many edge computing applications such as IIoT, robotics, autonomous driving, etc. Moreover, it's realized as the Eclipse Zenoh open-source project [3].

In this article, we'd like to provide an insight into the performance of all of the above pub/sub communication software and give a comparison between them, so that the developers may choose proper ones for the suitable applications which may benefit from them.

# The Software

In this section, Zenoh, MQTT, Kafka, and DDS will be briefly introduced.

## Zenoh

Eclipse zenoh unifies data in motion, data in use, data at rest, and computations. It carefully blends traditional pub/sub with geo-distributed storages, queries, and computations, while retaining a level of time and space efficiency that is well beyond any of the mainstream stacks.

With the implementation of the fully-decentralized architecture, Zenoh equips the applications with the capability of fault tolerance at the communication layer. This is different from most of the existing pub/sub communication systems which adopt centralized implementations. For example, both MQTT and Kafka need the role of brokers to bridge the communicating peers for data transmission. As a result, Zenoh won't have the issue of a single point of failure (SPOF) during communication between peers, while systems with centralized brokers may encounter some problems if the broker is broken. To reduce the chance of SPOF, centralized implementations typically need extra hardware/software resources such as multiple-machine clustering to increase the availability of the centralized brokers.

As a fully-distributed system, Zenoh supports peer-to-peer communication. Zenoh peers may create direct connections with each other for data communication whenever applicable. As a result, it may avoid the problem with communication bottlenecks, like that caused by the centralized brokers. Fig. 1 illustrates the models with a mixed topology. In the figure, Mesh and Clique are peer-to-peer models.

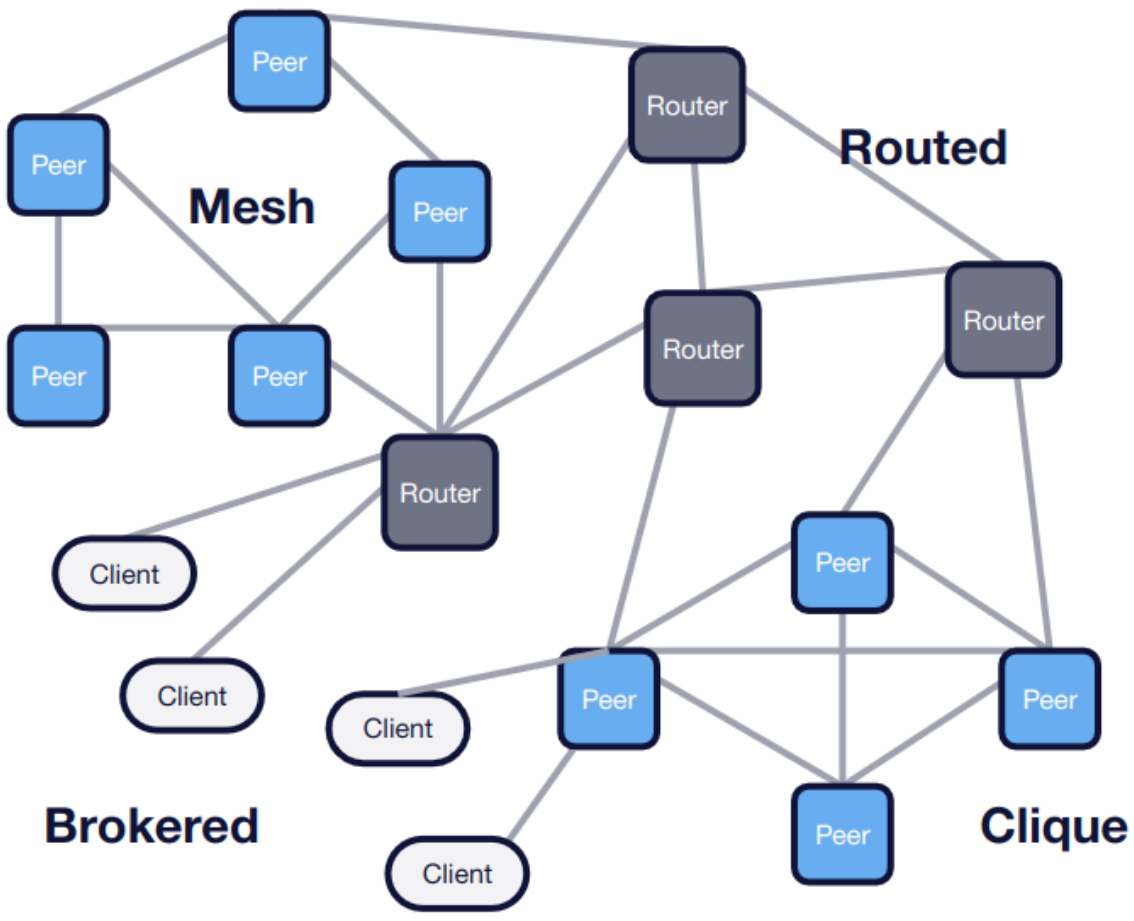

Fig. 1 Topology offered by Zenoh

In addition to the advantages from the availability and performance aspects, Zenoh also allows its network to be easily extended to the Internet scale by configuring itself across networks over the Internet, typically through the Zenoh routers for a suitable routing topology. Moreover, changing the Zenoh network topologies doesn't require changing the application logic and most of the Zenoh API for data transmission. Its connection configuration can be very flexible. The devices can be connected as mesh, clique, routed, or mixed topologies running under peer-to-peer or client mode depending on the scenarios.

Zenoh also allows different transport protocols to be used between peers, such as TCP, UDP, TLS/mTLS, QUIC, Websocket, UNIX-domain socket, and even non-IP protocols such as Bluetooth and serial data links [4]. Many optimization techniques have been incorporated to reduce the overheads and increase communication efficiency. And, Zenoh can be used in platforms from up to the cloud servers and down to the micro-controllers for sensors, supporting computation across the cloud-to-edge and to-things continuum.

## MQTT

MQTT is one of the most popular communication protocols used in IoT areas. It is designed as an extremely lightweight pub/sub messaging transport that is ideal for connecting remote devices with a small footprint and limited network bandwidth. The protocol defines two types of network entities: the message broker and clients (which include publishers and subscribers). The MQTT broker is a service that receives all messages from the publishers and then routes the messages to the appropriate destination (subscribers) according to the messages' topics.

The protocol provides three different QoS levels for communication. QoS level 0 represents the best-effort delivery of the message. QoS level 1 guarantees that a single message will be delivered to the destination clients at least once. And, QoS level 2 guarantees that a single message will be delivered to the destination clients exactly once.

## Kafka

Apache Kafka is a stream processing platform that aims to provide high throughput and low latency data feeds. The data are named by topics, which are similar to Zenoh, MQTT, and DDS. Producers may publish messages to different partitions of different topics, where messages will be ordered by offsets within each partition. Each consumer subscribing to one of the partitions keeps track of the offset consumed so far and pulls/consumes the data from the broker when the data has been published. The messages within a partition go in FIFO order. One producer can scatter the messages to different partitions to scale up the throughput. Kafka also supports replications, enabling large-scale data streaming applications.

Kafka provides nearly 700 options for brokers, topics, producers, and consumers. For example, the producer can control the number of records in a batch to send to the broker and the waiting time to send the next batch. The retention time and size of partitions for different topics can be tweaked. A broker can also be set to change the replication factor and the number of threads. These options give users the freedom to tune Kafka according to their needs.

In this article, we will use the terms producer/publisher and consumer/subscriber interchangeably for Kafka.

## DDS

DDS is an OMG standard. It has various implementations. In this article, we will use Eclipse Cyclone DDS [5] as the target DDS implementation to evaluate. Cyclone DDS is an open-source implementation of the OMG DDS specification, providing fast and robust data publication and subscription services in a network. It is one of the qualified Tier-1 RMW (ROS Middleware) implementations for ROS 2 [6] and is now shipped with ROS 2 [7] packages. Cyclone DDS already receives wide adoption in the robotics and autonomous vehicle industry and projects, such as TTTech Auto, Autoware, Apex.AI, AutoCore, and other adopters [8].

DDS leverages UDP multicast features to broadcast messages in the transport layer. It enables fast peer discovery and QoS-based low-latency data transmission within local networks, making it suitable for robotics applications and communications inside a vehicle. On the other hand, with UDP multicast by design, it's less suitable for communications over the open or wireless environment due to the flooding effect caused by the DDS discovery protocols. In addition, multicast is not well supported across routed networks. As a result, DDS itself is not a good choice for Internet communication. Zenoh, also developed by a team with rich experience in DDS implementation, resolves the issues by introducing advanced optimization and data-centric routing mechanisms in its protocol design.

# Experiment Method

Two scenarios were prepared for the experiments. The first ran the test programs on a single machine, and the other ran them on different machines connected through Ethernet. The averaged latency and throughput were then measured. Below lists the hardware configurations for both scenarios. Tab. 1 shows the configuration for the single machine. For the case of multiple machines, three of the same servers were used with a 100 Gb Ethernet connecting them together.

Tab. 1 The configuration of the testing machines

| | |
|---|---|
| OS | Ubuntu 20.04.3 LTS |
| CPU | AMD Ryzen 7 5800X running at a fixed frequency of 4.0 GHz<br>8-Core Processor, 16 threads |
| RAM | 32 GiB DDR4 3200MHz |
| Network | 100Gb Ethernet |

The relationship between the publisher, the subscriber, the broker, and the Zenoh router (or simply stated as the `router`) is illustrated in two diagrams in the subsections below for throughput and latency tests, respectively. For MQTT and Kafka, the messages must go through a broker. To provide a fair comparison, a Zenoh router was used in the Zenoh tests, and both the publisher and the subscriber were running with the client mode to form a routed topology in which all the data were relayed by the router to the destination. In addition, to further understand the performance that Zenoh can offer in different communication modes, the peer-to-peer mode (a.k.a. the peer mode) without having the data go through the Zenoh router was also added in the experiments as a reference. DDS also doesn't go through any broker and thus only supports the peer-to-peer mode.

For the single-machine scenario, all the programs were run in the same machine using `localhost` to specify the connections; while for the multiple-machine scenario, each of the publisher, the subscriber, and the broker/router were run on different machines. To perform the experiments, a set of scripts were created for measuring the throughput and latency on both the single-machine and the multiple-machine scenarios. The scripts were designed based on some suggestions from the guide for accurate measurement [9], and the testing environment was controlled by the scripts automatically. Note that for Zenoh's two test variations for the client mode and the peer mode, the same sets of test programs were used. Zenoh allows the communication mode to be easily changed by just setting the configuration file.

## Setup for Throughout Measurement

To measure the throughput, Zenoh, MQTT, Kafka, and Cyclone DDS were tested with payload sizes varying from 8 bytes (B) to 512  Mebibyte (MiB). In each measurement (or the term "sample" used in the test programs), the throughput is calculated whenever the subscriber has received a set of messages (with the number *N*) within a period of time (at least one second) to amortize the minor variation of the measurement. *N* is then divided by the elapsed time to get the number of messages per second (msg/s) as a metric. At least one test run is skipped to warm up the testing process, and the following measurements are recorded to calculate the average number.

The diagram for the throughput measurements is depicted below. For the tests with respect to the Zenoh client mode, MQTT, and Kafka, all the published data flow through the "Zenoh Router" or the "Broker" to the subscriber; while for the Zenoh peer mode and Cyclone DDS tests, the data will be passed from the publisher to the subscriber directly.

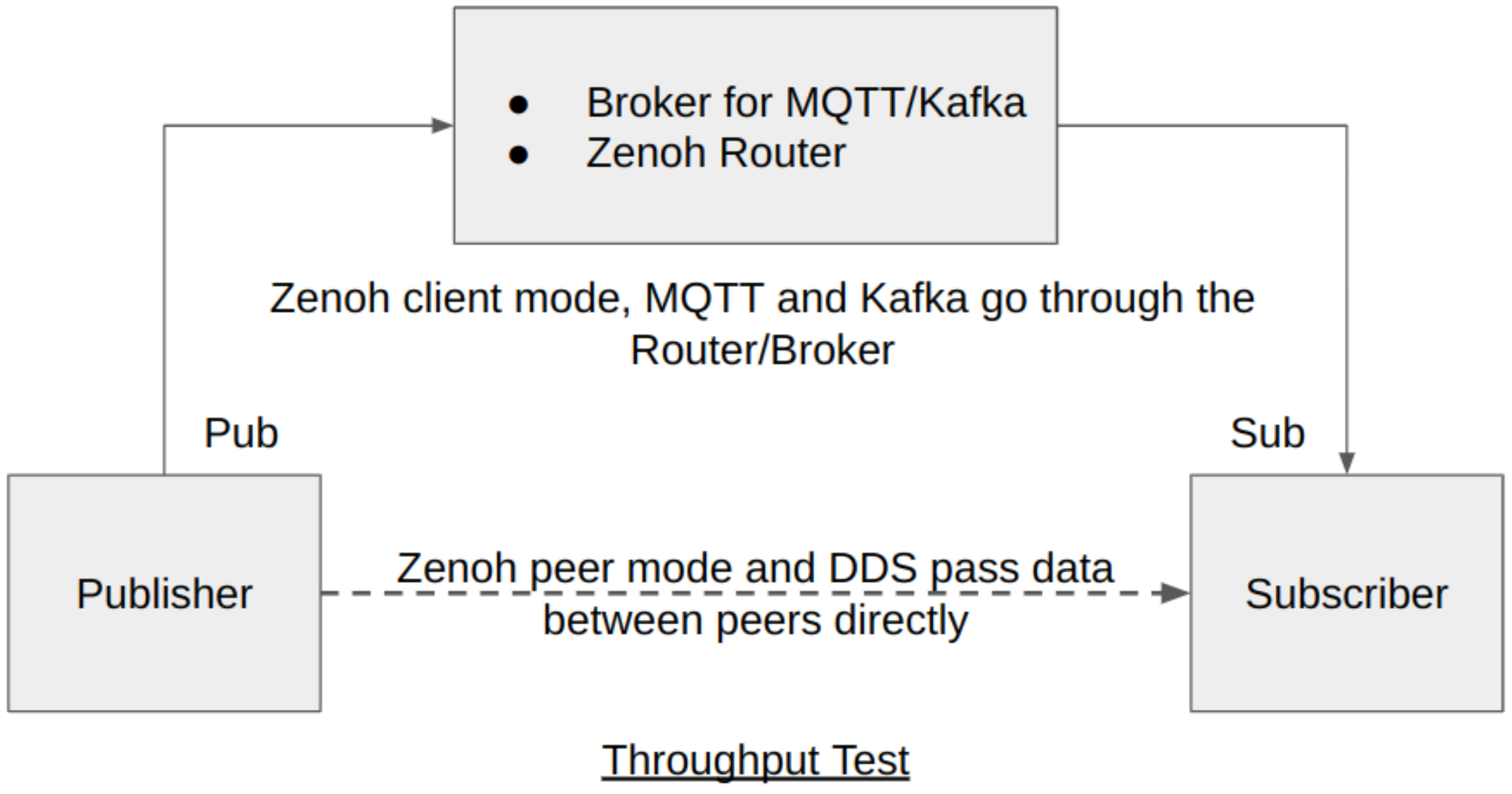

Fig. 2 The configuration diagram for throughput tests

## Setup for Latency Measurement

For latency, the ping-pong tests were used for the measurements. The numbered actions marked along the lines in the following figure indicate the order of the message transmission for one round of the ping-pong operation. In each run of measurement, the ping node sends a message (1 and 2 in the figure) and waits for it to be bounced back from the pong node (3 and 4). The tested payload size is fixed at 64 bytes (aligned with that for ICMP), and the testing is performed in a back-to-back manner to reduce the impact of the process scheduling and the context switches induced by the underlying operating system. The latency is defined to be half of the average round-trip time covering the ping and pong operations.

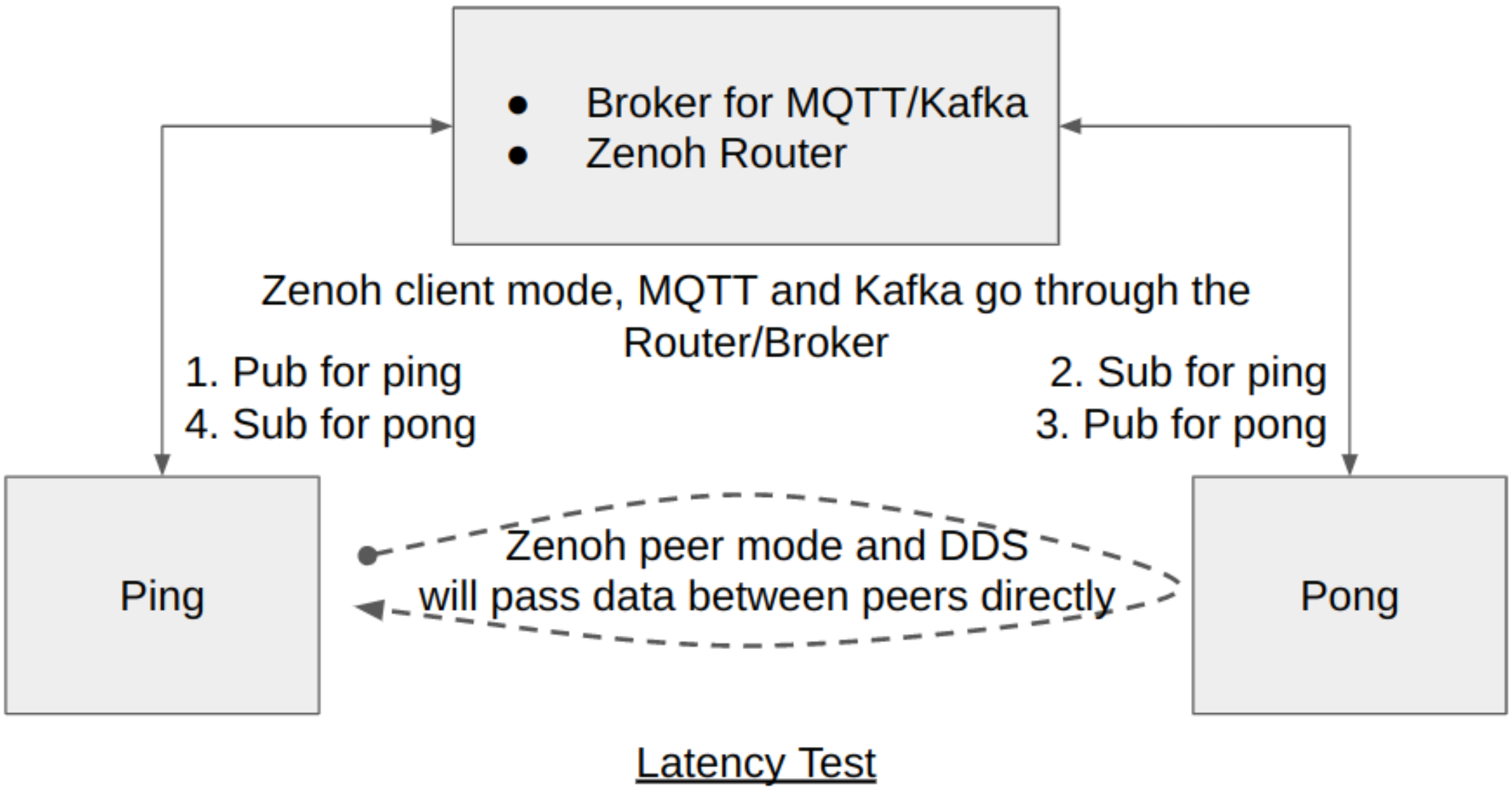

Fig. 3 The configuration diagram for latency tests

Similar to the case in the throughput test setup, all data of MQTT, Kafka, and the Zenoh client mode are relayed by the “Broker” or the “Zenoh Router”. On the other hand, the data for the Zenoh peer mode and Cyclone DDS can be passed between the ping and pong nodes directly.

## Option Setting for the Tests

The testing procedures are the same for Zenoh, Cyclone DDS, MQTT, and Kafka. All the benchmark programs can be found under the Zenoh performance test project [10]. Note that most of their options are kept with the default values, except for some minor options tuned specifically for MQTT and Kafka to get better performance for the measurements. The test conditions with the fine-tuned options are described in the following paragraphs.

The Zenoh test programs [11] use version 0.7.0-rc [12] and the Zenoh router `zenohd` can be built by following this guide [13]. For throughput testing, the publisher publishes data to the subscriber continuously. Its reliability was set to “reliable” and the congestion control option was set to "block" on the publisher side, which means when traffic jams occurred, the publisher would be blocked instead of dropping the messages. While the same congestion control was also used for the latency measurements, its reliability was set to “best effort” to align with the behavior of Kafka and MQTT. For all other options that can be tuned, the default settings were used.

The MQTT test programs [14] use Eclipse Mosquitto version 2.0.15 [15]. The MQTT clients were implemented with the Eclipse Paho MQTT C client library v.1.3.11 [16]. For all MQTT clients, the communication QoS level was set to 0 to achieve its best performance. For the broker’s settings, “max_inflight_messages” and “max_queued_messages” were both set to `unlimited` to reduce the negative impact on the performance according to the results of earlier tests. All other options remain as their default settings.

The Kafka test programs [17] use the official broker 3.2.1 [18] and the rdkafka client library 0.28.0 [19]. The configurations were tuned to achieve a higher message rate and lower latency for the benchmark programs. The following parameters were chosen for both throughput and latency test programs according to the suggestions from the online documents [20] and real experiments in order to get better results.

- linger.ms=0
- batch.size=400KB for throughput, 1 for latency
- compression.type=none
- acks=0
- fetch.min.bytes=1 (only for latency tests)

`linger.ms` allows the producer/publisher to wait for at most that number of milliseconds before sending each batch. Higher `linger.ms` implies longer latency. Although many articles suggest nonzero linger.ms for higher throughput, we see that linger.ms=0 was the best from real tests in our environment. `batch.size` causes the publisher to send the batch when the accumulated messages have reached the specified size. As the `batch.size` becomes larger, the to-be-published messages will wait longer. However, the data transmission overhead can be amortized in this way and the average overhead for each message will thus be reduced. Although the setting of linger.ms=0 may reduce the effect of the batch.size, a larger number of `batch.size=400KB` was still chosen in favor of higher throughput, while `batch.size=1B` was used in latency tests to make it perform better. Kafka allows the data to be compressed during transmission. To have a fair comparison with the other tested software and to shorten the latency, it's set to `none`.

From our experiments, `acks=0` benefited most on both throughput and latency tests. With `acks=1`, every time the publisher uploads a batch to the lead broker, the broker responds with an ack right away, while the default `acks=all` makes the broker respond with an ack after all replicas have received the record copies. In our environment, only one broker was used. To make Kafka work with the best performance, we chose to relax the criteria by setting `acks=0` so that the publisher does not wait for the response from the broker. For all other options, the default values were kept unchanged.

The DDS test programs use Eclipse Cyclone DDS [5] as the target implementation. They are based on the official DDS examples [22]. The following configuration was used by the examples.

- Reliability: RELIABLE, max blocking time 10 secs.
- One topic is used for each of the tests.
- Number of partitions is set to 1 for throughput and 2 for latency.

For the throughput test, the publisher and subscriber share the same partition for the topic. For the latency test, the ping and pong use different partitions under the same topic. The RELIABLE QoS is used to ensure that all data published can be delivered to the subscriber. In latency tests, the KEEP_LAST history QoS is used. In throughput tests, the KEEP_ALL history QoS was preferred over KEEP_LAST to further increase the reliability. Also, the subscribers take (by `dds_take()`) at most one sample per loop in both tests, in order to make the behavior analogous to Zenoh and others.

# Evaluation Results

The outcomes of throughput and latency are shown and discussed in this section. Each of these presents the results from single-machine and multiple-machine scenarios in order. In the following subsections, the lines specified by `Zenoh-brokered` in the figures are for Zenoh with the routed topology, and those with `Zenoh-p2p` are for the peer mode configuration.

## Throughput Results

In the throughput tests, the publisher program repeatedly published messages. The consumer program subscribed to the same topic collected a set of messages as a testing run and calculated the average message rate (msg/s). The same procedure applied to Zenoh, Cyclone DDS (briefly stated as DDS below), MQTT, and Kafka. The bitrates (bit/s) were also calculated accordingly based on the message rate and the payload size. In addition, the `iperf` utility was also used to measure the ideal data rate from the application layer between two peers on the same machines as a baseline reference. Various payload sizes were tested to see the trend of the throughput. For the charts in the following two subsections, the X-axis is for the payload size and the Y-axis is for the throughput in message rate and bitrate, respectively, notably in log scale.

### Single-machine Scenario

The throughput results for the single-machine scenario are first presented and explained. For message rates, Fig. 4 below shows the message rate for the single-machine scenario. The raw data can be found in Tab. 4 in the Appendix. The result show that Zenoh can reach up to more than 4M msg/s for small payload sizes. As the payload size increases, the throughput gradually decreases but is still better than those of the others. The maximum payload size measured in the experiment is 512 MiB. Among the two lines for Zenoh, the peer mode, indicated by the `Zenoh-p2p` line, provides better results than that of the client mode, marked by `Zenoh-brokered`, because the data transmission doesn’t need to be relayed and thus limited by the Zenoh router. DDS, although slower, is relatively close to Zenoh and can still reach up to 2M msg/s. The trend of the message rate for DDS is also quite aligned with that for Zenoh with respect to the payload size.

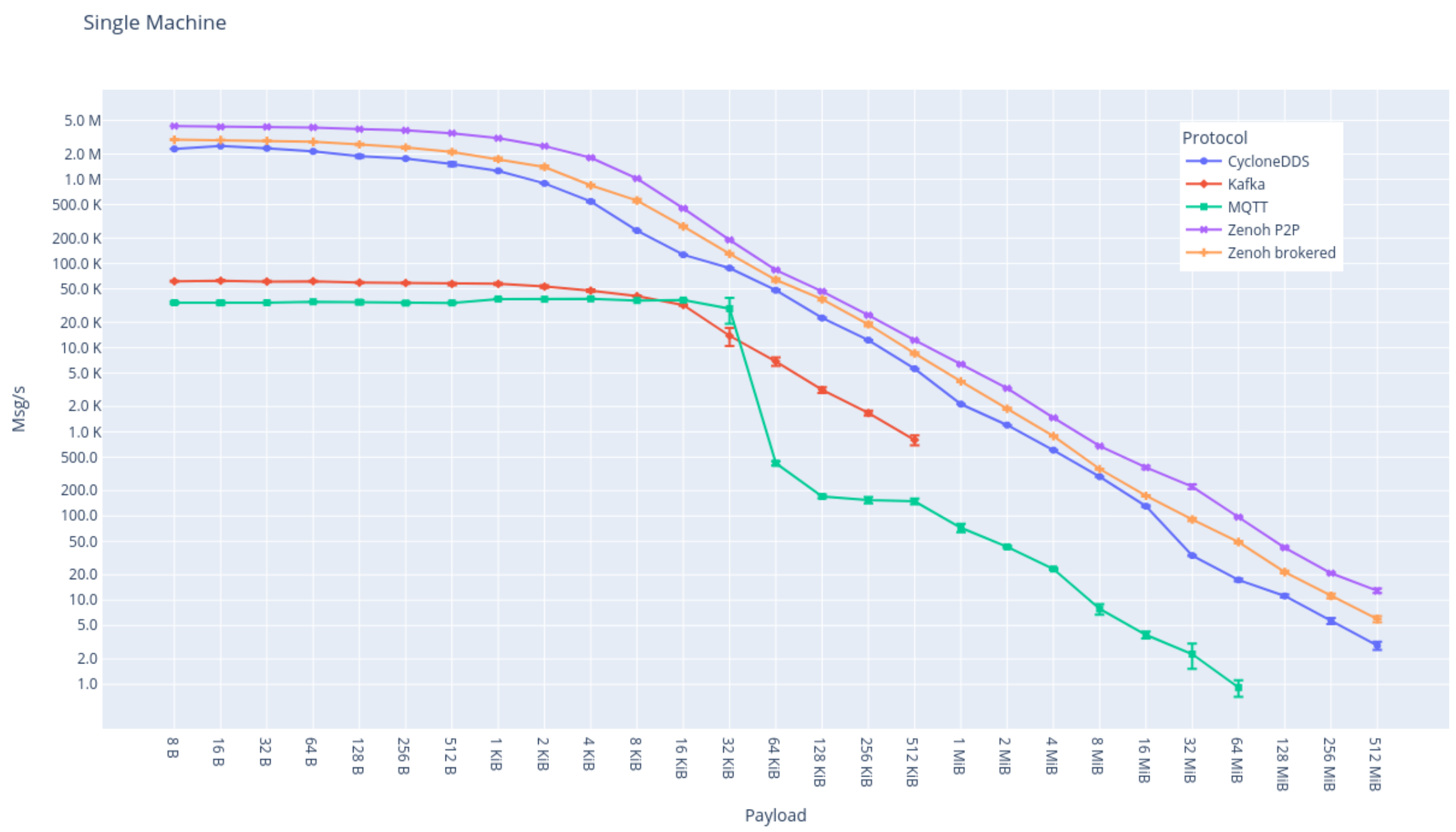

Fig. 4 Throughput data in msg/s for the single-machine scenario

Kafka, from the same figure, maintains a stable message rate between 53K to 62K msg/s for the payload sizes less than 2 KiB and starts to decrease until the last data that was successfully measured at the payload size of 512 KiB. MQTT has a slightly lower throughput between 32K to 38K msg/s before 32 KiB of the payload size but starts to reduce drastically after that and shows the worst performance number among all the software. From the results, MQTT appears to support better than Kafka regarding the successfully transferred payloads. However, Kafka in principle should be able to support more payload sizes. The reason behind the observed limitation is that the Kafka bindings we used failed on extending the message size over 1 MB.

As a summary for this, Zenoh with the peer mode can be 2x the message rate than DDS, and over 60x and 100x better than Kafka and MQTT at the payload size of 8 bytes, respectively. And, it still retains 2x, 6x, and 12x message rates compared to DDS, MQTT, and Kafka around the closest points at the payload size of 32 KiB.

Fig. 5 is for the throughput numbers from the viewpoint of bits-per-second (bit/s or bps). The corresponding raw data can be found in Tab. 5 in the Appendix. It shows that Zenoh starts to saturate close to the ideal throughput measured by `iperf` (drawn in a light-blue-dashed line at 76 Gpbs) for the payload size equal to or larger than 4 KiB. The throughput obtained by `iperf` here reflects the best number that can be achieved with the given network software stack on the target machine. Zenoh can reach up to 37 Gbps for the configuration of the client mode shown in `Zenoh-brokered` and can be further increased to 67 Gbps for the peer mode configuration shown in `Zenoh-p2p`.

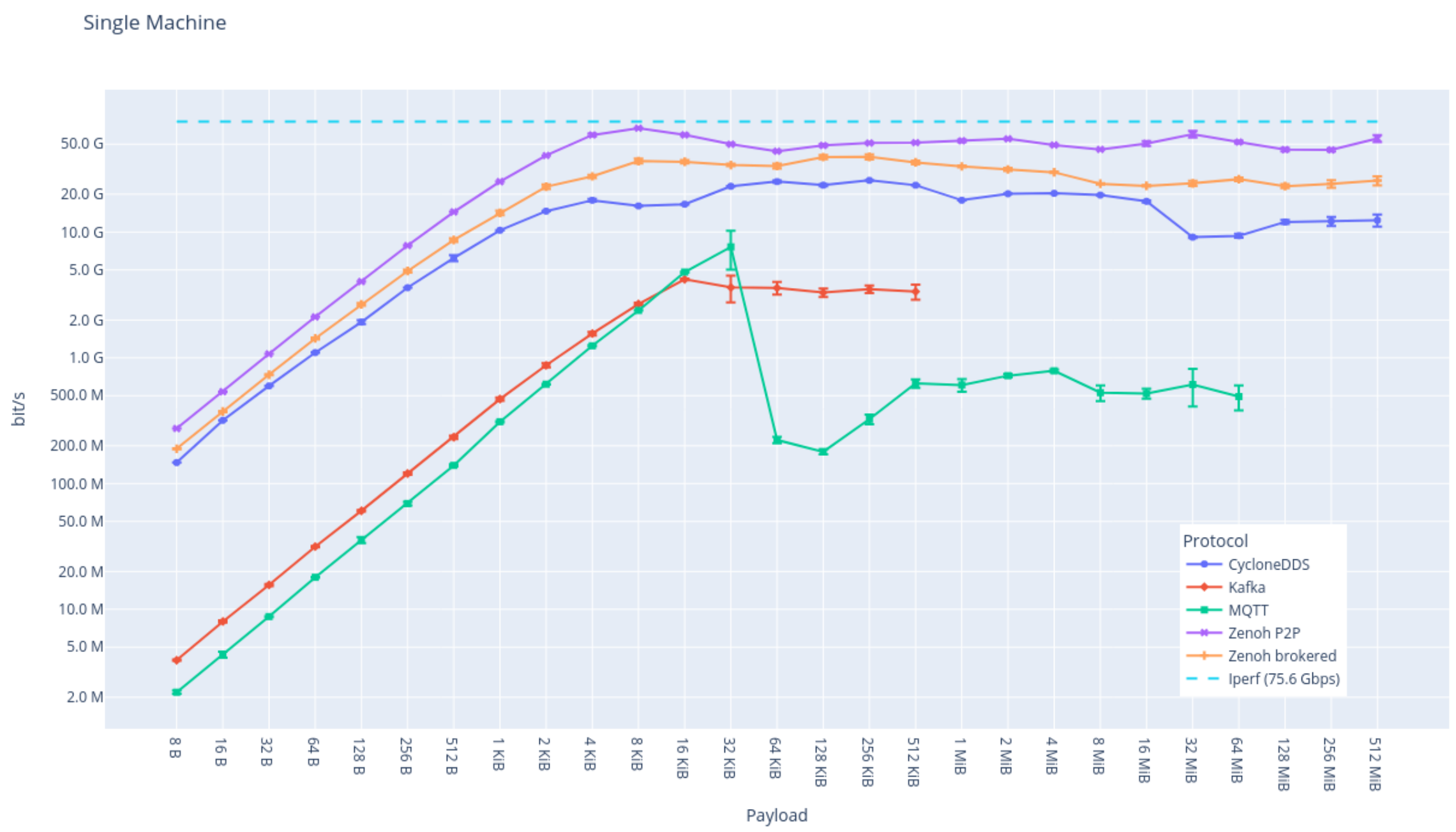

Fig. 5 Throughput data in bit/s for the single-machine scenario

For DDS, the throughput is lower than Zenoh (with the highest one as 26 Gpbs) but still much higher than Kafka and MQTT. Kafka appears to saturate at about 3~4 Gbps when the payload size is larger than 16 KiB but fails to work when the payload size is larger than 512 KiB. For MQTT, it reaches up to 8 Gbps at the payload size of 32 KiB but then starts to decrease greatly as mentioned previously. The performance degradation phenomenon of MQTT appeared in the tests consistently. The reason behind this is intriguing and will be worth further investigation in the future.

Overall, in this single-machine scenario, taking bitrate numbers for the throughput comparison, considering the best numbers obtained from Zenoh, DDS, Kafka, and MQTT, Zenoh peer mode achieves ~2x higher performance compared to that of DDS and 65x and 130x for Kafka and MQTT at the payload size of 8 bytes. Zenoh achieves peak performances at 8KB, achieving more than 4x throughput than DDS, 22x than Kafka, and 33x than MQTT. Finally, for a payload of 32KB  Zenoh achieves more than 2x higher throughput than DDS, 6x than MQTT, and 12x than Kafka.

## Multiple-machine Scenario

To evaluate the performance over a physical network, the same tests were also performed on multiple machines. In our tests, different machines were used for each of the publisher, the subscriber, and the broker/Zenoh-router, respectively, with 100 Gb Ethernet. The throughput results for various payload sizes in such a scenario are shown in the Fig. 6 and 7 below for the message rate (msg/s) and the data rate (bit/s), respectively. The corresponding raw data can be found in Tab. 6 and 7 in the Appendix.

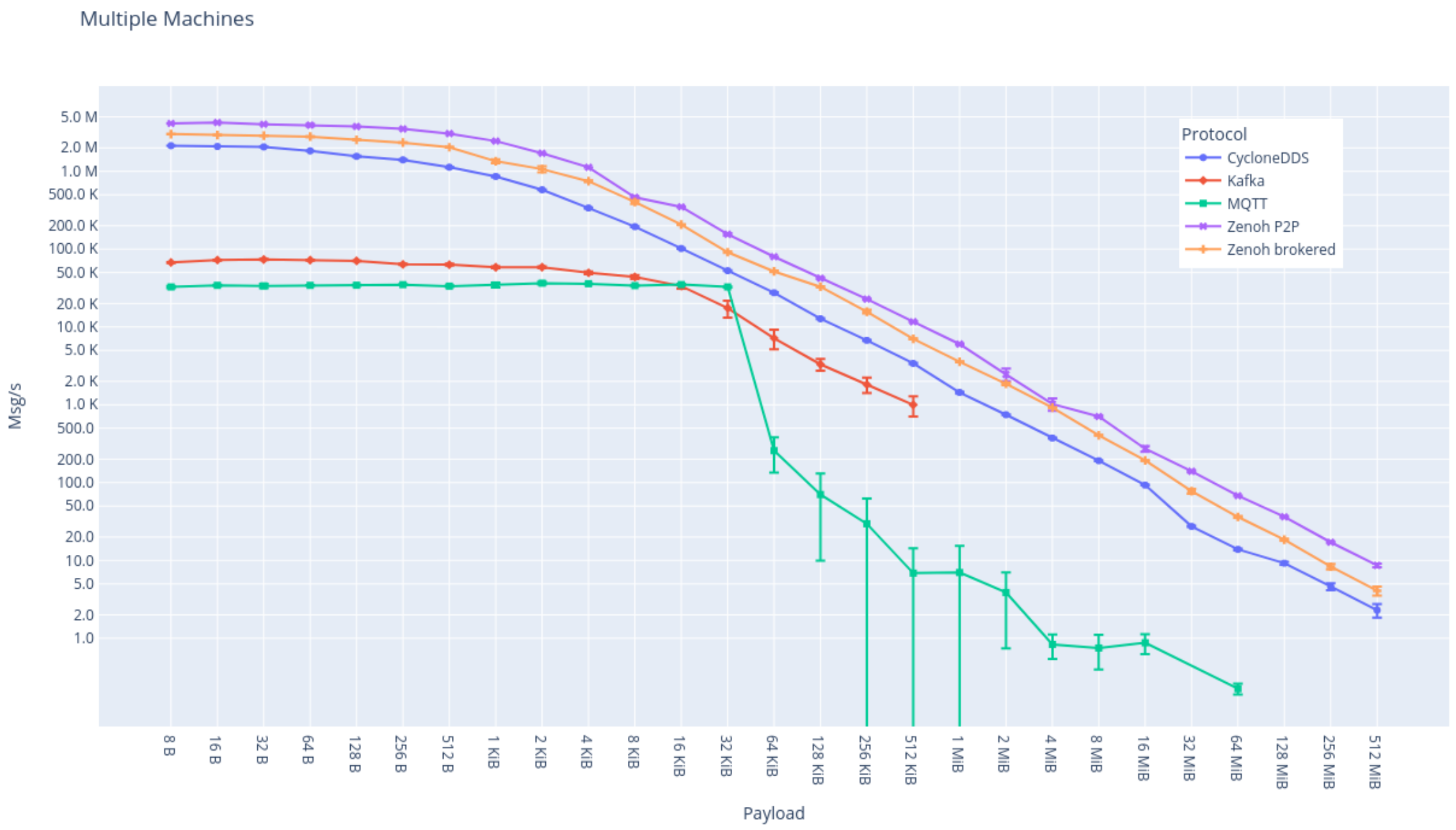

Fig. 6 Throughput data in msg/s for the multiple-machine scenario

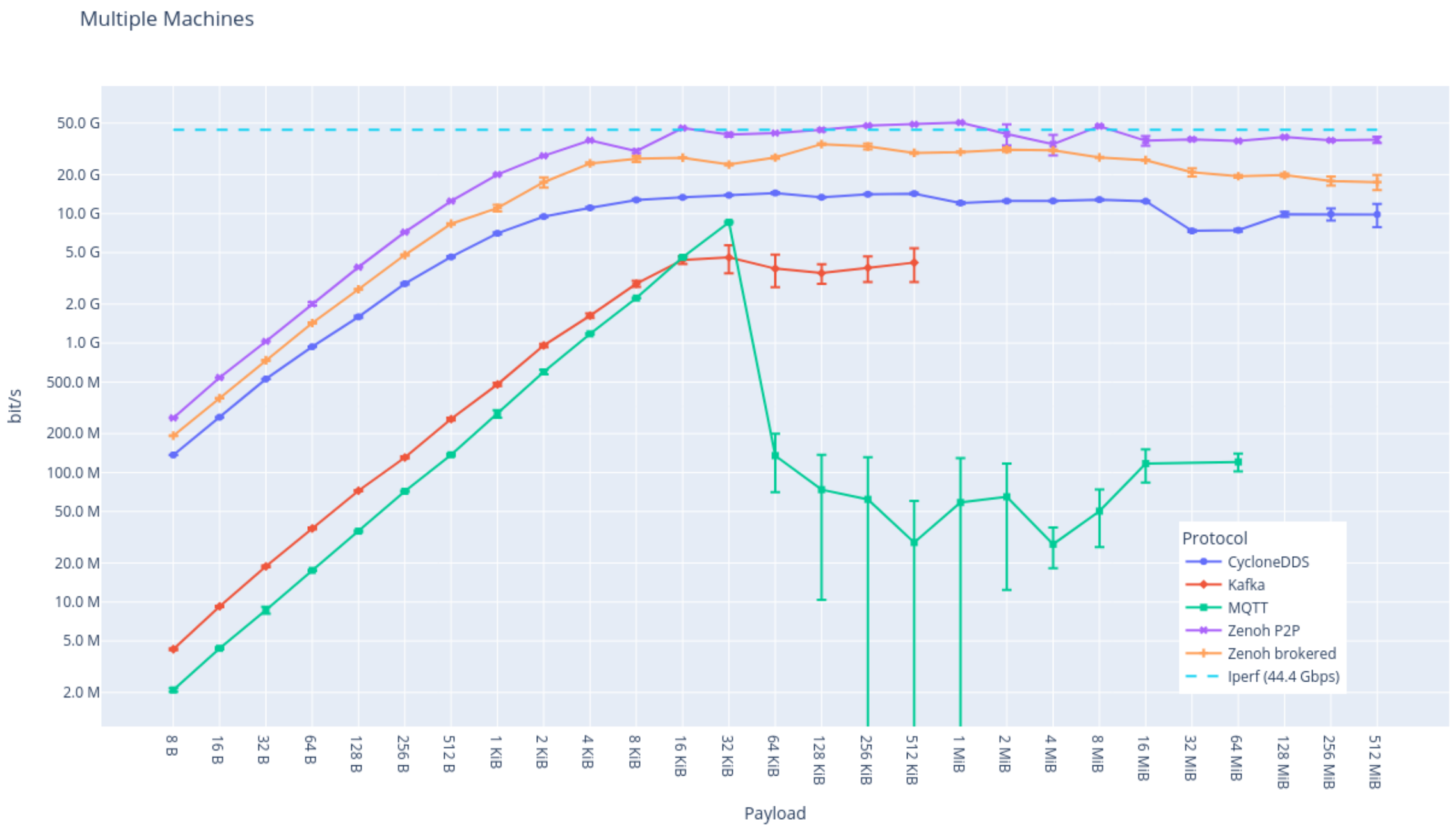

Fig. 7 Throughput data in bit/s for the multiple-machine scenario

As the results of the message rate and the bitrate are correlated and the behavior shown here is similar to that of the single-machine scenario, we will mainly focus on discussing the bitrate results at this place. In the figures, the light-blue-dashed line at 44 Gpbs for `iperf` shows the ideal throughput of the target network to provide a baseline reference for the testing environment with a physical network. From the results, we can see that Zenoh's throughput still outperforms DDS, Kafka, and MQTT for all payload sizes, with a similar trend observed in the single-machine scenario.

As expected, Zenoh with the peer mode is better than that with the client mode. The maximal bitrate is about 34 Gbps for `Zenoh-brokered` and 51 Gbps for `Zenoh-p2p`, respectively. Note that the peer mode doesn't require the data to be relayed by the router and thus the number of messages per pub/sub pair is cut in half. For the peer mode, its performance is still relatively aligned with the ideal throughput obtained by `iperf`, when the payload size is larger than 2 KiB.

For Cyclone DDS, its throughput ranking remains at number three in the charts. The maximum bitrate is 14 Gbps. On the other hand, MQTT's bitrates reach up to 9 Gbps at the payload size of 32 KiB, quite close to the results of the single-machine scenario, mainly due to its simple and lightweight design. However, the results after that are worse. It goes down with an even higher slope. For Kafka, its best bitrate number is 5 Gbps, also at the payload size of 32 KiB. It actually performs better than the single-machine scenario at payload sizes smaller than 2 KiB, because the processing load of the broker was separated from the producer and consumer. Notably, MQTT and Kafka failed the tests for payload sizes larger than 64 MiB and 512 KiB, respectively.

As a summary of the comparison covering the message rate and the bitrate, from the figures, at the payload size of 8 bytes, Zenoh with the peer mode can achieve 2x, 55x, and 110x throughput compared to DDS, Kafka, and MQTT, respectively; and for the other reference point at payload size of 32 KiB, Zenoh can still be over 2x, 4x, and 8x better than that of DDS, MQTT, and Kafka, respectively.

# Latency Results

In the latency tests, the `ping` program publishes the ping message, and the `pong` program replies with the same message upon receiving the ping. The round-trip time (RTT) was measured for each ping/pong pair and then the latency value was calculated as RTT/2. As mentioned previously, the tested payload size is fixed at 64 bytes and the ping-pong testing was performed as fast as possible to approach the ideal latency value with less impact from the OS scheduling.

## Single-machine Scenario

Tab. 2 shows the results of the tests inside the single-machine environment. It reflects the overheads for the measured software. The Linux `ping` utility was included to measure the minimum latency which can be achieved with the network stack inside the machine (1 us).

Tab. 2 Latency data in microseconds for the single-machine scenario

| Target | Latency (us) |
|---|---|
| Kafka | 73 |
| MQTT | 27 |
| Cyclone DDS | 8 |
| Zenoh-brokered | 21 |
| Zenoh-p2p | 10 |
| Zenoh-pico | 5 |
| ping | 1 |

Among the tested programs, MQTT and Kafka have a latency of 73 us and 27 us, respectively. As for Zenoh, while `Zenoh-brokered` has a latency of 21 us, similar to but lower than that of MQTT, mainly due to the similar behavior of routing data through a middleman (the Zenoh router or the broker for MQTT), `Zenoh-p2p` shows that it can be further reduced to 10 us.

For Cyclone DDS, the latency is even lower than Zenoh, achieving down to 8 us. The reason is that it's using UDP Multicast. Although Zenoh currently hasn't implemented the same data transport yet, the microcontroller implementation of Zenoh – Zenoh-pico – has already realized this. As a result, we've also tested its latency and the result is 5 us, as indicated by `Zenoh-pico`, which is even lower than that of Cyclone DDS, mainly because the Zenoh protocol and its implementation can be more lightweight and efficient.

In general, the latency of Zenoh is promising. Zenoh with the default peer-to-peer mode has the shortest latency compared to MQTT and Kafka. When the UDP Multicast transport is supported by Zenoh, as the result indicated by Zenoh-pico, it is expected that Zenoh can achieve the lowest number among all the testees including Cyclone DDS, and approach the ideal value represented by `ping`.

## Multiple-machine Scenario

For the latency numbers over a real network with 100 Gb Ethernet, the results are depicted in the following table.

Tab. 3 Latency data in microseconds for the multiple-machine scenario

| Target | Latency (us) |
|---|---|
| Kafka | 84 |
| MQTT | 45 |
| Cyclone DDS | 38 |
| Zenoh-brokered | 41 |
| Zenoh-p2p | 16 |
| Zenoh-pico | 13 |
| ping | 7 |

For Zenoh using the client mode with the Zenoh router, the latency is about 41 us, which is higher than the results from the single-machine scenario because the data needs to be passed to and forwarded by the router to the destination over the physical network; that is, the data is transmitted over the wire four times, as depicted in Fig. 3. When such a condition is removed by allowing the data to be transferred from the publisher to the subscriber directly, the overhead can be reduced to a minimum, as indicated by the number of `Zenoh-p2p` for the peer mode as 16 us, which is much lower than that for MQTT (45 us) and Kafka (84 us). While Cyclone DDS now has a longer latency at 38 us, even higher than `Zenoh-p2p`, `Zenoh-pico` remains the best one at 13 us, closest to the baseline obtained by the `ping` utility at 7 us.

# Conclusion

In this article, we compare the performance of Zenoh, MQTT, Kafka, and Cyclone DDS. The throughput and latency were measured by the basic pub-sub and the ping-pong operations, respectively.

Two scenarios were evaluated. The first one was performed on a single machine to see the software overhead, by assuming an almost-unlimited bandwidth inside the same machine. The second scenario was performed on multiple machines, in which the evaluated peers, i.e. the publisher/subscriber and the ping/pong program pairs, ran on separate machines connected to a physical network. Although Zenoh and DDS support direct peer-to-peer communication between the communicating peers, for the sake of a fair comparison with the centralized broker configuration used by MQTT and Kafka, a set of Zenoh tests was also configured to use the client mode which passes all data through the Zenoh router.

From the comparison results, we can see that for throughput, Zenoh consistently outperformed MQTT, Kafka, and DDS, given similar experimental conditions. Zenoh achieved up to 67 Gbps on single-machine tests and 51 Gbps on multiple-machine tests with a 100 GbE network when the peer mode was used. For the largest performance gaps in bitrates, Zenoh even shows one to two orders of magnitude improvement on MQTT and Kafka (more than 50x and 100x in both scenarios) and could double the throughput of Cyclone DDS. For the closest numbers, Zenoh can still be several times the performance compared to the others.

For latency, Zenoh has lower numbers than MQTT and Kafka. Although DDS obtains shorter latency than Zenoh does for the single-machine scenario due to its use of the UDP multicast transport mechanism, we can see that the microcontroller implementation of Zenoh, i.e. Zenoh-pico, which has already realized the UDP multicast transport, can perform better than DDS in all cases. This capability will be supported by all Zenoh implementations in the future. For the multiple-machine scenario, both Zenoh with the peer mode and Zenoh-pico appear to be better than DDS, showing its efficiency on a physical link.

For Zenoh itself, the peer mode performed better than the client mode, because the former allows the data to be transmitted between the communicating peers directly without going through a middleman, i.e. the broker or the Zenoh router.

From the results shown in this article, we can see that Zenoh is relatively closer to the ideal numbers obtained by the classic baseline tools `iperf` and `ping` for throughput and latency evaluation from the application layer, thanks to the low overhead design and multiple optimization techniques embedded in Zenoh's implementation. In addition to the performance difference, Zenoh is the only one that was able to successfully and stably deliver messages with a continuously growing payload size.

In conclusion, MQTT, Kafka, and DDS are well and long-established technologies with a broader user base and many real applications in their respective fields. Zenoh, on the other hand, is a newly designed protocol implemented by a team with rich experience in distributed systems and DDS. The goal of Zenoh is to provide another option for the industry, especially for the environment with time-sensitive connectivity, reliability, and scalability requirements.

From the result of the performance comparison, we can see that Zenoh generally outperformed MQTT, Kafka, and DDS for the basic throughput and latency benchmarks. We believe Zenoh will be one of the best choices for industrial and IoT applications that can seamlessly support the cloud-to-edge and to-things continuum.

# Acknowledgments

Special thanks to **Carlos Guimarães** and **Gabriele Baldoni** for helping run the tests in the ideal experiment environment, and **Luca Cominardi**, **Angelo Corsaro**, **Olivier Hecart,** and the whole **Zenoh Team in Paris** for helping review and solve problems. Also thanks to **Eddie Wang** and **Rex Yu**, former interns of the Zenoh Taiwan Team, for helping at the beginning of this evaluation.

# Appendix

Tab. 4 Raw Data for Single-machine Throughput (msg/s)

| Payload size | CycloneDDS | Kafka | MQTT | Zenoh P2P | Zenoh brokered |
|---|---|---|---|---|---|
| 8 B | 2.3 M ± 1.2 K | 61.6 K ± 724.5 | 34.2 K ± 1.0 K | 4.3 M ± 7.4 K | 3.0 M ± 28.3 K |
| 16 B | 2.5 M ± 3.1 K | 62.5 K ± 1.3 K | 34.1 K ± 1.9 K | 4.2 M ± 2.2 K | 2.9 M ± 16.4 K |
| 32 B | 2.3 M ± 7.7 K | 61.1 K ± 1.2 K | 34.2 K ± 292.1 | 4.2 M ± 4.4 K | 2.9 M ± 28.3 K |
| 64 B | 2.1 M ± 5.4 K | 61.6 K ± 618.0 | 35.1 K ± 185.1 | 4.1 M ± 2.8 K | 2.8 M ± 42.0 K |
| 128 B | 1.9 M ± 68.1 K | 59.4 K ± 809.5 | 34.8 K ± 1.8 K | 3.9 M ± 7.4 K | 2.6 M ± 61.5 K |
| 256 B | 1.8 M ± 3.3 K | 58.7 K ± 1.2 K | 33.9 K ± 1.1 K | 3.8 M ± 23.3 K | 2.4 M ± 68.7 K |
| 512 B | 1.5 M ± 85.9 K | 57.1 K ± 1.7 K | 34.1 K ± 661.1 | 3.5 M ± 2.2 K | 2.1 M ± 68.5 K |
| 1 KiB | 1.3 M ± 4.5 K | 57.3 K ± 1.6 K | 37.9 K ± 401.9 | 3.1 M ± 7.1 K | 1.7 M ± 69.6 K |
| 2 KiB | 895.5 K ± 2.0 K | 53.2 K ± 2.0 K | 37.7 K ± 236.1 | 2.5 M ± 3.8 K | 1.4 M ± 55.8 K |
| 4 KiB | 546.4 K ± 12.7 K | 47.6 K ± 1.5 K | 38.1 K ± 396.0 | 1.8 M ± 45.5 K | 846.6 K ± 25.1 K |
| 8 KiB | 246.2 K ± 700.7 | 40.9 K ± 832.5 | 36.3 K ± 103.8 | 1.0 M ± 7.0 K | 560.1 K ± 24.7 K |
| 16 KiB | 127.1 K ± 693.1 | 32.1 K ± 474.7 | 36.7 K ± 311.1 | 452.0 K ± 7.5 K | 276.3 K ± 8.3 K |
| 32 KiB | 88.2 K ± 559.0 | 13.8 K ± 3.3 K | 29.1 K ± 9.9 K | 190.6 K ± 2.9 K | 130.4 K ± 2.9 K |
| 64 KiB | 48.1 K ± 1.0 K | 6.9 K ± 785.6 | 423.9 ± 24.8 | 83.8 K ± 1.2 K | 63.9 K ± 2.9 K |
| 128 KiB | 22.5 K ± 439.4 | 3.2 K ± 244.2 | 170.8 ± 8.0 | 46.6 K ± 1.1 K | 37.7 K ± 1.5 K |
| 256 KiB | 12.3 K ± 33.5 | 1.7 K ± 114.6 | 155.0 ± 13.4 | 24.3 K ± 291.0 | 18.9 K ± 903.2 |
| 512 KiB | 5.6 K ± 16.6 | 800.9 ± 108.6 | 149.4 ± 11.5 | 12.3 K ± 112.0 | 8.5 K ± 276.9 |
| 1 MiB | 2.1 K ± 2.8 | N/A | 72.5 ± 8.3 | 6.3 K ± 139.2 | 4.0 K ± 66.3 |
| 2 MiB | 1.2 K ± 3.2 | N/A | 43.0 ± 1.3 | 3.3 K ± 29.3 | 1.9 K ± 53.1 |
| 4 MiB | 606.4 ± 0.7 | N/A | 23.5 ± 0.5 | 1.5 K ± 24.7 | 890.7 ± 14.2 |
| 8 MiB | 293.5 ± 0.7 | N/A | 7.9 ± 1.1 | 676.7 ± 1.2 | 361.1 ± 3.7 |
| 16 MiB | 130.9 ± 2.6 | N/A | 3.9 ± 0.4 | 377.1 ± 16.2 | 173.8 ± 2.7 |
| 32 MiB | 33.9 ± 0.5 | N/A | 2.3 ± 0.8 | 223.8 ± 13.2 | 90.9 ± 4.1 |
| 64 MiB | 17.4 ± 0.6 | N/A | 0.9 ± 0.2 | 96.9 ± 2.0 | 49.1 ± 1.6 |
| 128 MiB | 11.2 ± 0.4 | N/A | N/A | 42.1 ± 1.3 | 21.6 ± 0.8 |
| 256 MiB | 5.7 ± 0.5 | N/A | N/A | 20.9 ± 0.6 | 11.2 ± 0.8 |
| 512 MiB | 2.9 ± 0.3 | N/A | N/A | 12.9 ± 0.8 | 6.0 ± 0.5 |

Tab. 5 Raw Data for Single-machine Throughput (bps)

| Payload size | CycloneDDS | Kafka | MQTT | Zenoh P2P | Zenoh brokered |
|---|---|---|---|---|---|
| 8 B | 147.0 M ± 76.3 K | 3.9 M ± 46.4 K | 2.2 M ± 67.1 K | 274.2 M ± 475.0 K | 189.4 M ± 1.8 M |
| 16 B | 318.6 M ± 396.4 K | 8.0 M ± 164.5 K | 4.4 M ± 244.2 K | 537.5 M ± 275.4 K | 373.9 M ± 2.1 M |
| 32 B | 598.0 M ± 2.0 M | 15.6 M ± 308.2 K | 8.8 M ± 74.8 K | 1.1 G ± 1.1 M | 734.2 M ± 7.2 M |
| 64 B | 1.1 G ± 2.8 M | 31.5 M ± 316.4 K | 18.0 M ± 94.8 K | 2.1 G ± 1.4 M | 1.4 G ± 21.5 M |
| 128 B | 1.9 G ± 69.7 M | 60.8 M ± 828.9 K | 35.6 M ± 1.9 M | 4.0 G ± 7.6 M | 2.7 G ± 63.0 M |
| 256 B | 3.6 G ± 6.9 M | 120.3 M ± 2.5 M | 69.4 M ± 2.3 M | 7.8 G ± 47.6 M | 4.9 G ± 140.6 M |
| 512 B | 6.2 G ± 351.9 M | 233.9 M ± 7.0 M | 139.6 M ± 2.7 M | 14.5 G ± 9.0 M | 8.7 G ± 280.7 M |
| 1 KiB | 10.3 G ± 36.5 M | 469.7 M ± 13.0 M | 310.5 M ± 3.3 M | 25.2 G ± 58.5 M | 14.2 G ± 570.4 M |
| 2 KiB | 14.7 G ± 33.4 M | 871.7 M ± 32.9 M | 617.1 M ± 3.9 M | 40.6 G ± 61.5 M | 23.0 G ± 913.8 M |
| 4 KiB | 17.9 G ± 414.8 M | 1.6 G ± 50.7 M | 1.2 G ± 13.0 M | 59.2 G ± 1.5 G | 27.7 G ± 823.7 M |
| 8 KiB | 16.1 G ± 45.9 M | 2.7 G ± 54.6 M | 2.4 G ± 6.8 M | 67.0 G ± 458.0 M | 36.7 G ± 1.6 G |
| 16 KiB | 16.7 G ± 90.8 M | 4.2 G ± 62.2 M | 4.8 G ± 40.8 M | 59.2 G ± 979.9 M | 36.2 G ± 1.1 G |
| 32 KiB | 23.1 G ± 146.5 M | 3.6 G ± 876.3 M | 7.6 G ± 2.6 G | 50.0 G ± 771.9 M | 34.2 G ± 752.6 M |
| 64 KiB | 25.2 G ± 535.0 M | 3.6 G ± 411.9 M | 222.2 M ± 13.0 M | 44.0 G ± 620.8 M | 33.5 G ± 1.5 G |
| 128 KiB | 23.6 G ± 460.7 M | 3.3 G ± 256.1 M | 179.1 M ± 8.4 M | 48.9 G ± 1.1 G | 39.5 G ± 1.5 G |
| 256 KiB | 25.8 G ± 70.2 M | 3.5 G ± 240.4 M | 325.0 M ± 28.1 M | 51.0 G ± 610.3 M | 39.7 G ± 1.9 G |
| 512 KiB | 23.6 G ± 69.7 M | 3.4 G ± 455.4 M | 626.5 M ± 48.2 M | 51.5 G ± 469.8 M | 35.8 G ± 1.2 G |
| 1 MiB | 17.9 G ± 23.7 M | N/A | 608.1 M ± 69.6 M | 53.2 G ± 1.2 G | 33.3 G ± 556.1 M |
| 2 MiB | 20.2 G ± 53.2 M | N/A | 721.4 M ± 21.7 M | 55.1 G ± 491.0 M | 31.5 G ± 890.6 M |
| 4 MiB | 20.3 G ± 23.1 M | N/A | 788.5 M ± 17.9 M | 49.2 G ± 827.7 M | 29.9 G ± 475.7 M |
| 8 MiB | 19.7 G ± 49.1 M | N/A | 528.4 M ± 75.5 M | 45.4 G ± 82.6 M | 24.2 G ± 250.3 M |
| 16 MiB | 17.6 G ± 345.3 M | N/A | 520.1 M ± 47.5 M | 50.6 G ± 2.2 G | 23.3 G ± 357.7 M |
| 32 MiB | 9.1 G ± 125.7 M | N/A | 613.6 M ± 202.9 M | 60.1 G ± 3.5 G | 24.4 G ± 1.1 G |
| 64 MiB | 9.3 G ± 321.6 M | N/A | 492.1 M ± 109.6 M | 52.0 G ± 1.1 G | 26.3 G ± 876.2 M |
| 128 MiB | 12.0 G ± 448.1 M | N/A | N/A | 45.2 G ± 1.4 G | 23.2 G ± 856.8 M |
| 256 MiB | 12.2 G ± 1.0 G | N/A | N/A | 45.0 G ± 1.3 G | 24.1 G ± 1.6 G |
| 512 MiB | 12.4 G ± 1.4 G | N/A | N/A | 55.5 G ± 3.4 G | 25.7 G ± 2.1 G |

Tab. 6 Raw Data for Multiple-machine Throughput (msg/s)

| Payload size | CycloneDDS | Kafka | MQTT | Zenoh P2P | Zenoh brokered |
|---|---|---|---|---|---|
| 8 B | 2.1 M ± 5.8 K | 67.4 K ± 959.3 | 32.7 K ± 1.1 K | 4.1 M ± 8.6 K | 3.0 M ± 9.6 K |
| 16 B | 2.1 M ± 3.8 K | 72.3 K ± 782.2 | 34.3 K ± 636.1 | 4.2 M ± 2.5 K | 2.9 M ± 3.1 K |
| 32 B | 2.1 M ± 8.0 K | 73.6 K ± 1.2 K | 33.8 K ± 2.1 K | 4.0 M ± 8.1 K | 2.9 M ± 18.8 K |
| 64 B | 1.8 M ± 6.0 K | 72.3 K ± 1.2 K | 34.3 K ± 274.4 | 3.9 M ± 123.0 K | 2.8 M ± 21.5 K |
| 128 B | 1.6 M ± 26.4 K | 70.7 K ± 757.6 | 34.4 K ± 485.6 | 3.8 M ± 15.0 K | 2.5 M ± 2.4 K |
| 256 B | 1.4 M ± 19.5 K | 63.8 K ± 1.2 K | 35.0 K ± 400.0 | 3.5 M ± 10.2 K | 2.3 M ± 21.4 K |
| 512 B | 1.1 M ± 15.3 K | 63.3 K ± 1.6 K | 33.5 K ± 654.2 | 3.1 M ± 10.3 K | 2.0 M ± 25.1 K |
| 1 KiB | 860.1 K ± 17.1 K | 58.5 K ± 1.6 K | 34.7 K ± 2.4 K | 2.4 M ± 29.4 K | 1.3 M ± 78.7 K |
| 2 KiB | 579.6 K ± 5.5 K | 58.5 K ± 1.8 K | 36.6 K ± 1.5 K | 1.7 M ± 2.7 K | 1.1 M ± 97.0 K |
| 4 KiB | 338.4 K ± 4.7 K | 49.7 K ± 2.1 K | 36.0 K ± 340.2 | 1.1 M ± 1.3 K | 745.1 K ± 16.2 K |
| 8 KiB | 194.6 K ± 2.5 K | 43.9 K ± 2.4 K | 34.0 K ± 201.3 | 463.4 K ± 13.8 K | 404.3 K ± 21.6 K |
| 16 KiB | 102.0 K ± 878.8 | 33.5 K ± 2.3 K | 35.1 K ± 554.7 | 349.6 K ± 3.4 K | 206.3 K ± 712.8 |
| 32 KiB | 53.0 K ± 554.5 | 17.5 K ± 4.3 K | 32.7 K ± 425.1 | 155.3 K ± 5.6 K | 91.4 K ± 1.3 K |
| 64 KiB | 27.5 K ± 457.0 | 7.2 K ± 2.0 K | 257.6 ± 123.1 | 79.9 K ± 100.3 | 51.8 K ± 1.2 K |
| 128 KiB | 12.8 K ± 153.4 | 3.3 K ± 572.3 | 70.4 ± 60.5 | 42.3 K ± 938.9 | 32.8 K ± 430.8 |
| 256 KiB | 6.7 K ± 88.6 | 1.8 K ± 408.3 | 29.6 ± 33.0 | 22.8 K ± 286.4 | 15.8 K ± 825.7 |
| 512 KiB | 3.4 K ± 49.6 | 998.2 ± 290.4 | 6.9 ± 7.5 | 11.7 K ± 148.6 | 7.0 K ± 116.7 |
| 1 MiB | 1.4 K ± 33.5 | N/A | 7.0 ± 8.4 | 6.0 K ± 78.3 | 3.6 K ± 42.8 |
| 2 MiB | 745.9 ± 7.7 | N/A | 3.9 ± 3.1 | 2.5 K ± 451.4 | 1.9 K ± 69.7 |
| 4 MiB | 373.7 ± 3.8 | N/A | 0.8 ± 0.3 | 1.0 K ± 187.4 | 919.9 ± 7.6 |
| 8 MiB | 191.2 ± 2.0 | N/A | 0.8 ± 0.4 | 707.3 ± 15.9 | 405.4 ± 3.6 |
| 16 MiB | 92.9 ± 0.8 | N/A | 0.9 ± 0.2 | 272.6 ± 23.3 | 193.0 ± 1.7 |
| 32 MiB | 27.4 ± 0.5 | N/A | N/A | 139.8 ± 3.6 | 77.7 ± 5.6 |
| 64 MiB | 13.9 ± 0.3 | N/A | 0.2 ± 0.0 | 67.8 ± 0.4 | 36.2 ± 0.9 |
| 128 MiB | 9.2 ± 0.4 | N/A | N/A | 36.3 ± 0.6 | 18.5 ± 0.5 |
| 256 MiB | 4.6 ± 0.5 | N/A | N/A | 17.1 ± 0.3 | 8.3 ± 0.7 |
| 512 MiB | 2.3 ± 0.5 | N/A | N/A | 8.6 ± 0.5 | 4.1 ± 0.5 |

Tab. 7 Raw Data for Multiple-machine Throughput (bps)

| Payload size | CycloneDDS | Kafka | MQTT | Zenoh P2P | Zenoh brokered |
|---|---|---|---|---|---|
| 8 B | 136.6 M ± 371.2 K | 4.3 M ± 61.4 K | 2.1 M ± 68.0 K | 264.1 M ± 548.2 K | 192.4 M ± 617.1 K |
| 16 B | 267.8 M ± 482.7 K | 9.3 M ± 100.1 K | 4.4 M ± 81.4 K | 541.5 M ± 320.6 K | 374.9 M ± 399.6 K |
| 32 B | 527.9 M ± 2.0 M | 18.9 M ± 305.6 K | 8.7 M ± 539.2 K | 1.0 G ± 2.1 M | 735.7 M ± 4.8 M |
| 64 B | 937.8 M ± 3.1 M | 37.0 M ± 595.8 K | 17.5 M ± 140.5 K | 2.0 G ± 63.0 M | 1.4 G ± 11.0 M |
| 128 B | 1.6 G ± 27.0 M | 72.4 M ± 775.8 K | 35.3 M ± 497.2 K | 3.9 G ± 15.3 M | 2.6 G ± 2.5 M |
| 256 B | 2.9 G ± 39.9 M | 130.7 M ± 2.4 M | 71.7 M ± 819.2 K | 7.2 G ± 20.9 M | 4.8 G ± 43.8 M |
| 512 B | 4.6 G ± 62.6 M | 259.2 M ± 6.5 M | 137.2 M ± 2.7 M | 12.5 G ± 42.2 M | 8.4 G ± 102.8 M |
| 1 KiB | 7.0 G ± 140.1 M | 479.5 M ± 13.5 M | 284.1 M ± 19.5 M | 20.1 G ± 241.1 M | 11.0 G ± 645.0 M |
| 2 KiB | 9.5 G ± 90.3 M | 959.0 M ± 29.0 M | 599.2 M ± 24.5 M | 28.0 G ± 45.0 M | 17.5 G ± 1.6 G |
| 4 KiB | 11.1 G ± 153.1 M | 1.6 G ± 67.5 M | 1.2 G ± 11.1 M | 36.9 G ± 41.1 M | 24.4 G ± 531.0 M |
| 8 KiB | 12.8 G ± 166.5 M | 2.9 G ± 155.7 M | 2.2 G ± 13.2 M | 30.4 G ± 902.2 M | 26.5 G ± 1.4 G |
| 16 KiB | 13.4 G ± 115.2 M | 4.4 G ± 302.4 M | 4.6 G ± 72.7 M | 45.8 G ± 443.8 M | 27.0 G ± 93.4 M |
| 32 KiB | 13.9 G ± 145.4 M | 4.6 G ± 1.1 G | 8.6 G ± 111.4 M | 40.7 G ± 1.5 G | 24.0 G ± 343.5 M |
| 64 KiB | 14.4 G ± 239.6 M | 3.8 G ± 1.1 G | 135.1 M ± 64.5 M | 41.9 G ± 52.6 M | 27.1 G ± 620.8 M |
| 128 KiB | 13.4 G ± 160.9 M | 3.5 G ± 600.1 M | 73.8 M ± 63.4 M | 44.4 G ± 984.5 M | 34.4 G ± 451.7 M |
| 256 KiB | 14.1 G ± 185.9 M | 3.8 G ± 856.3 M | 62.1 M ± 69.1 M | 47.8 G ± 600.6 M | 33.0 G ± 1.7 G |
| 512 KiB | 14.3 G ± 208.0 M | 4.2 G ± 1.2 G | 28.8 M ± 31.5 M | 49.0 G ± 623.4 M | 29.5 G ± 489.5 M |
| 1 MiB | 12.1 G ± 281.3 M | N/A | 58.7 M ± 70.9 M | 50.5 G ± 656.5 M | 29.9 G ± 359.1 M |
| 2 MiB | 12.5 G ± 129.4 M | N/A | 65.0 M ± 52.6 M | 41.3 G ± 7.6 G | 31.3 G ± 1.2 G |
| 4 MiB | 12.5 G ± 126.0 M | N/A | 28.0 M ± 9.7 M | 34.4 G ± 6.3 G | 30.9 G ± 254.2 M |
| 8 MiB | 12.8 G ± 132.8 M | N/A | 50.3 M ± 23.7 M | 47.5 G ± 1.1 G | 27.2 G ± 238.6 M |
| 16 MiB | 12.5 G ± 113.9 M | N/A | 117.4 M ± 33.6 M | 36.6 G ± 3.1 G | 25.9 G ± 231.4 M |
| 32 MiB | 7.4 G ± 133.6 M | N/A | N/A | 37.5 G ± 975.0 M | 20.9 G ± 1.5 G |
| 64 MiB | 7.4 G ± 186.1 M | N/A | 120.8 M ± 19.0 M | 36.4 G ± 237.3 M | 19.5 G ± 480.3 M |
| 128 MiB | 9.9 G ± 448.1 M | N/A | N/A | 39.0 G ± 622.8 M | 19.8 G ± 544.1 M |
| 256 MiB | 9.9 G ± 1.1 G | N/A | N/A | 36.8 G ± 744.3 M | 17.9 G ± 1.4 G |
| 512 MiB | 9.9 G ± 2.0 G | N/A | N/A | 37.1 G ± 2.1 G | 17.5 G ± 2.3 G |